\documentclass[reprint,superscriptaddress,amsmath,amsfonts,amssymb,aps,prl,showkeys]{revtex4-1}
\usepackage{hyperref}
\usepackage{graphicx}
\usepackage{xcolor}
\hypersetup{
  breaklinks = {true},
  citecolor = {blue},
  colorlinks = {true},
  linkcolor = {red},
}
\usepackage{soul}

\begin{document}

%\title{Spin-orbit torques in CrTe\textsubscript{2} and Janus CrXTe (X=S,Se) for an all-in-one spin-orbit torque platform}
\title{Giant Spin-Orbit Torque in Cr-based Janus Transition Metal Dichalcogenides} 

\author{Libor Voj\'{a}\v{c}ek}
\thanks{Authors contributed equally}
\affiliation{Universit\'{e} Grenoble Alpes, CEA, CNRS, IRIG-Spintec, 38000 Grenoble, France}

\author{Joaqu{\'i}n Medina Due{\~n}as}
\thanks{Authors contributed equally}
\affiliation{ICN2 --- Catalan Institute of Nanoscience and Nanotechnology, CSIC and BIST, Campus UAB, Bellaterra, 08193 Barcelona, Spain}
\affiliation{Department of Physics, Universitat Aut{\`o}noma de Barcelona (UAB), Campus UAB, Bellaterra, 08193 Barcelona, Spain}

\author{Jing Li}
\affiliation{Universit\'{e} Grenoble Alpes, CEA, Leti, F-38054, Grenoble, France}

\author{Fatima Ibrahim}
\affiliation{Universit\'{e} Grenoble Alpes, CEA, CNRS, IRIG-Spintec, 38000 Grenoble, France}

\author{Aurélien Manchon}
\affiliation{Aix-Marseille Université, CNRS, CINAM, Marseille 13288, France}

\author{Stephan Roche}
\affiliation{ICN2 --- Catalan Institute of Nanoscience and Nanotechnology, CSIC and BIST,
Campus UAB, Bellaterra, 08193 Barcelona, Spain}
\affiliation{ICREA --- Instituci\'o Catalana de Recerca i Estudis Avan\c{c}ats, 08010 Barcelona, Spain}

\author{Mairbek Chshiev}
\affiliation{Universit\'{e} Grenoble Alpes, CEA, CNRS, IRIG-Spintec, 38000 Grenoble, France}
\affiliation{Institut Universitaire de France, 75231 Paris, France}

\author{Jos{\'e} H. Garc{\'i}a}
\affiliation{ICN2 --- Catalan Institute of Nanoscience and Nanotechnology, CSIC and BIST, Campus UAB, Bellaterra, 08193 Barcelona, Spain}

\date{\today}

\begin{abstract}
We report a very large spin-orbit torque (SOT) capability of chromium-based transition metal dichalcogenides (TMD) in their Janus forms CrXTe, with X=S,Se. 
The structural inversion symmetry breaking, inherent to Janus structures is responsible for a large SOT response generated by giant Rashba splitting, equivalent to that obtained by applying a transverse electric field of $\sim 100 \,\text{V} \,\text{nm}^{-1}$ in non-Janus CrTe\textsubscript{2}, completely out of experimental reach. By performing transport simulations on custom-made Wannier tight-binding models, Janus systems are found to exhibit a SOT performance comparable to the most efficient two-dimensional materials, while allowing for field-free perpendicular magnetization switching owing to their reduced in-plane symmetry.
Altogether, our findings evidence that magnetic Janus TMDs stand as suitable candidates for ultimate SOT-MRAM devices.

% \JMD{OLD: }We report very large intrinsic spin-orbit torque (SOT) capability of magnetic Janus two-dimensional materials based on 1T-phase Transition Metal Dichalcogenides (TMDs) such as CrXTe with X=S,Se. Using first-principles approaches and custom-made Wannier tight-binding models providing accurate spin textures, we perform quantum transport simulations of chromium-based Janus TMD SOT components using the Kubo-Bastin formalism.  The breaking of inversion symmetry inherent to Janus CrXTe is found to produce a giant torque figure of merit, which would correspond to the transverse electric field of up to $\sim 100 \,\text{V} \,\text{nm}^{-1}$ in non-Janus CrTe\textsubscript{2} materials, completely out of experimental reach. Our findings evidence that magnetic Janus TMD stand as suitable candidates for ultimate SOT-MRAM.
\end{abstract}

\maketitle

\textit{Introduction}.---
The spin-orbit torque (SOT) mechanism represents an innovative method to electrically manipulate the magnetization of a magnetic material \cite{manchon_current-induced_2019, tian_two-dimensional_2021}, providing remarkable energy-efficiency, writing speed and scalability prospects which have earned their insertion in magnetic random access memory (MRAM) applications \cite{guo_spintronics_2021, yang_two-dimensional_2022}, among other developing technologies \cite{dieny_opportunities_2020, han_spin-orbit_2021, fert_electrical_2024}.
SOT-MRAM prototype cells have been shown to operate on the sub-nanosecond timescale \cite{garello_ultrafast_2014, garello_spin-orbit_2019, jhuria_spin-orbit_2020, cubukcu_ultra-fast_2018}, with a power consumption of merely one percent of their (already in commercial use) spin-transfer torque counterpart \cite{garello_SOT-MRAM_2018, song_high_2022}.
Although next-generation SOT-based technologies advance at a steady pace, they still face issues regarding massive density and integration. The development of SOT-MRAMs remains limited to multi-layered devices, where SOT figures of merit are strongly sensitive to interface quality; while additionally, a densely packed SOT-MRAM requires electrical switching of a perpendicular magnetic anisotropy (PMA)~\cite{dieny_perpendicular_2017}, which is only achieved in conventional devices with the assistance of an external magnetic field. 

Two-dimensional materials offer alternative paths to overcome these issues. Atomically clean interfaces have been shown to enhance both charge-to-spin conversion \cite{husain_large_2020, bonell_control_2020, hidding_spin-orbit_2020} and tunneling magnetoresistance \cite{karpan_graphite_2007, wang_tunneling_2018} while reducing the cell dimensions. Furthermore, precise control of crystal symmetries enables novel SOT mechanisms that allow for \textit{field-free} PMA switching \cite{macneill_control_2017, liu_symmetry-dependent_2021}.
Materials such as CrTe\textsubscript{2} \cite{freitas_ferromagnetism_2015, sun_room_2020, zhang_room-temperature_2021, liu_structural_2022} and Fe$_n$GeTe\textsubscript{2} ($n=3, 4, 5$) \cite{fei_two-dimensional_2018, deng_gate-tunable_2018, seo_nearly_2020, may_ferromagnetism_2019, ribeiro_large-scale_2022} offer the most promising alternatives to overcome the usually low Curie temperature ($T_\text{C}$) of van der Waals ferromagnets (FM); where additional gating, strain and chemical composition engineering allow to tune their magnetic properties \cite{deng_gate-tunable_2018, may_tuning_2020, 
tian_tunable_2020, wang_strain-sensitive_2020}.
New avenues for SOT devices are opened when exploiting the metallic nature of these materials along with their strong spin-orbit coupling (SOC), overcoming multi-layer designs in favor of an all-in-one platform where the FM acts as both the SOC material and the free magnetization in a self-induced SOT scheme \cite{kurebayashi_antidamping_2014, tang_bulk_2020, liu_electrical_2020,Hidding_2021,hidding_role_2023}. In this context, Janus transition metal dichalcogenide (TMD) monolayers stand out as the material of choice \cite{liang_very_2020, smaili_janus_2021, zhang_two-dimensional_2021}. Indeed, the CrXTe ultrathin layers, similar to their non-Janus counterpart CrTe\textsubscript{2}~\cite{meng_anomalous_2021}, are expected to be magnetic with PMA under the adequate experimental conditions, high $T_\text{C}$ even exceding room temperature, and are most stable in their metallic 1T-phase \cite{cui_strain-tunable_2020, liu_enhanced_2023}. Inversion symmetry forbids a SOT response in CrTe\textsubscript{2}; however, this stepback is overcome by breaking the symmetry between both chalcogen atoms, for instance, by applying an electric field transversal to the crystal plane. Another symmetry-breaking mechanism is obtained by substituting one Te atom with S or Se, thus forming the Janus CrXTe structures, where the symmetry breaking is manifested in the crystal field. 

In this Letter, using first principles approach and quantum transport simulations performed on Wannier tight-binding models carefully designed to represent the reciprocal spin textures, together with critical field-free PMA switching current calculations, we report an exceptional SOT performance of chromium-based Janus TMD monolayers CrXTe, with X=S,Se. We compare both the Janus and non-Janus materials under electric field, demonstrating that Cr-based Janus monolayers constitute an optimal SOT platform for low-energy magnetization reversal.

\begin{figure*}[t!]
    \centering
    \includegraphics[width=\textwidth]{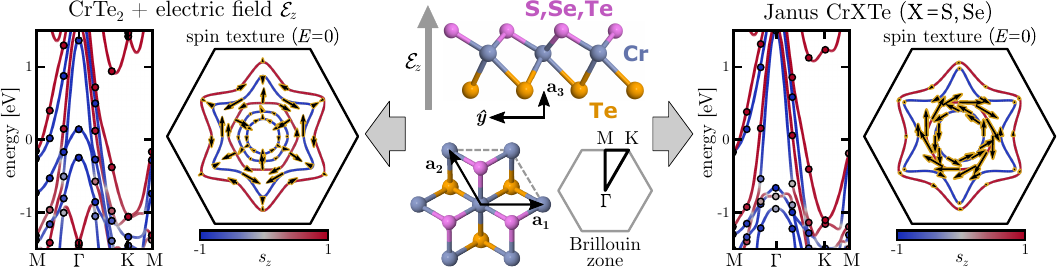}
    \caption{Middle Panel: CrXTe Janus structure, lattice vectors and Brillouin zone. Left panel: Band structure and spin-textures of symmetric CrTe\textsubscript{2} under the effect of an applied electric field [$\mathcal{E}=2 \,\text{V} \,\text{nm}^{-1}$]. The continuous band structure curves and black spin texture arrows are obtained from the Wannier model, while the discrete band structure dots and yellow highlighted spin texture arrows are obtained from DFT calculations. Right panel: Band structure and spin textures of CrXTe [X=Se].}
    \label{fig:fig1}
\end{figure*}

\textit{Electronic structure and spin textures}.--- 
We perform first-principles calculations of the selected TMDs using density functional theory (DFT)~\cite{kresse_ab_1993, kresse_efficiency_1996} with GGA-PBE pseudopotentials~\cite{perdew_generalized_1996-1} and an effective Hubbard~\textit{U} correction of 3.0~eV to localize the Cr-d orbitals. The usual metallic 1T phase displayed at the center of Fig.~ \ref{fig:fig1} proves to be more stable than the semiconducting 1H phase. Based on the maximally-localized Wannier functions~\cite{marzari_maximally_2012, mostofi_updated_2014}, we then derive tight-binding models of the DFT ground state. Accurate representations require a 22-orbital basis comprising the Cr-d and the chalcogen-p bands, with the interactions expanded into a 25$\times$25 supercell. The \textit{ab initio} band structure is thereby represented with an impressive $\sim 1 \,\text{meV}$ accuracy in a large window of $\approx -5$ to $+4$~eV around the Fermi level. We also carefully derive the real-space spin operator in the Wannier basis, resulting in a non-local spin that replicates the \textit{ab initio} spin texture with $\sim 1$\% error \cite{supmat}.

Overall, the comparison between DFT results and those of the Wannier model yields an excellent agreement, as shown in Fig.~\ref{fig:fig1}. This demonstrates the capability of our methodology, allowing to reach a DFT-level accuracy via tight-binding models for general systems.

Note that the remarkable SOT capability of the low-symmetry Janus systems is already apparent from its Fermi-level spin textures, showing a prominent helical winding in CrXTe, displayed in the right panel of Fig.~\ref{fig:fig1}. This large Rashba term comes from a strong inversion-symmetry breaking due to a huge \textit{internal} out-of-plane electric field $\sim $ 1--2 $\text{V} \,\text{nm}^{-1}$ caused by a charge imbalance at the two inequivalent chalcogen atoms S/Se and Te~\cite{supmat}. On the contrary, an \textit{external} electric field $\mathcal{E}_z$ applied on CrTe\textsubscript{2} to break its innate inversion symmetry is heavily screened due to its large dielectric constant $\varepsilon_\mathrm{CrTe2}$, reducing any applied field by a factor of $\varepsilon_\mathrm{CrTe2}^{-1} \sim 0.02$ inside the layer~\cite{supmat}. This is apparent in the left panel of Fig.~\ref{fig:fig1} where the odd-in-momentum Rashba helical winding induced by the (heavily screened) external field is very small compared to the visibly prominent even-in-momentum spin texture stemming from the CrTe\textsubscript{2} centrosymmetric crystal field. Achieving a Rashba term in CrTe\textsubscript{2} similar to that of the Janus CrXTe would require an immense applied field $\mathcal{E}_z \sim \varepsilon_\mathrm{CrTe2} \cdot$1--2 $\text{V} \,\text{nm}^{-1} \approx$ 50--100 $\text{V} \,\text{nm}^{-1}$~\cite{supmat}. This illustrates the remarkable SOT potential of Janus CrXTe already at this ground-state level.

\textit{Non-equilibrium response}.---
CrXTe generally exhibits $C_{3v}$ (also denoted 3m) point group symmetry. In the case of CrTe\textsubscript{2}, the degeneracy between both chalcogens raises the symmetry group to $D_{3d}$, which includes inversion symmetry and thus forbids any SOT response. However, $C_{3v}$ symmetry is recovered by applying an out-of-plane electric field. The magnetization dynamics, described by the Landau-Lifshitz-Gilbert equation, is driven by the torque surface density $\boldsymbol{T}$, which is determined by the non-equilibrium spin surface density $\boldsymbol{S}$ as 
\begin{equation}
\boldsymbol{T} = \hbar^{-1} J_\text{ex} \hat{\boldsymbol{m}} \times \boldsymbol{S},
\end{equation}
where $\hat{\boldsymbol{m}}$ is the magnetization direction unit vector, and $J_\text{ex}$ is the exchange energy which couples the localized magnetic moments with those of the itinerant electrons \cite{manchon_current-induced_2019}. The exact form of the non-equilibrium response is dictated by the subjacent crystal symmetries via invariant theory \cite{garcia-ovalle_spin-orbit_2023}. The non-equilibrium spin density, expanded up to first order with respect to both the driving electric field $\boldsymbol{\mathcal{E}}$ and the magnetization direction $\hat{\boldsymbol{m}}$, reads
\begin{equation}
\begin{split}
    \boldsymbol{S} =& \,\chi_\text{FL}^{} \hat{\boldsymbol{z}} \times \boldsymbol{\mathcal{E}} - \chi_\text{DL}^{} \hat{\boldsymbol{m}} \times (\hat{\boldsymbol{z}} \times \boldsymbol{\mathcal{E}}) - \chi_\text{DL}^{z} (\hat{\boldsymbol{m}} \cdot \boldsymbol{\mathcal{E}}) \hat{\boldsymbol{z}} \\& + \chi_\text{3m}^{} [(\mathcal{E}_x m_y + \mathcal{E}_y m_x) \hat{\boldsymbol{x}} + (\mathcal{E}_x m_x - \mathcal{E}_y m_y) \hat{\boldsymbol{y}}] \text{ ,}
\end{split}
\end{equation}
with $\chi_\alpha^{}$ the spin linear response coefficients, and $\tau_\alpha^{} = \hbar^{-1} J_\text{ex} \chi_\alpha^{}$ the associated spin-torque conductivity. 
The terms proportional to $\chi_\text{FL}^{}$ and $\chi_\text{DL}^{}$ represent the conventional field-like and damping-like torques respectively, while $\chi_\text{DL}^{z}$ corresponds to an out-of-plane anisotropy of the damping-like torque. These torques are general to arbitrary non-centrosymmetric systems and drive the magnetization to an \textit{in-plane} stationary state along the $\hat{\boldsymbol{z}} \times \boldsymbol{\mathcal{E}}$ direction. Thus, additional fields are required in order to switch a system with \textit{perpendicular} magnetic anisotropy. The term proportional to $\chi_\text{3m}^{}$ represents the so-called 3m torque, which is particular to systems with 3m symmetry. This contribution is related to a current-induced in-plane magnetic anisotropy \cite{johansen_current_2019} which modifies the stationary state of the magnetization inducing an out-of-plane instability, thus enabling field-free switching of a perpendicular magnetization \cite{liu_symmetry-dependent_2021}. We remit the reader to the supplementary material for a more complete calculation of the non-equilibrium spin density \cite{supmat}.

Within the Kubo quantum transport framework, we calculate the non-equilibrium spin density at Fermi level $\varepsilon_\text{F}^{}$ using the Kubo-Bastin formula
\begin{equation}
    \boldsymbol{S}(\varepsilon_\text{F}^{}) = -2\hbar \int \text{d}\varepsilon \mathit{f}(\varepsilon) \text{Im} \, \text{Tr} \left[ \delta(\varepsilon - \hat{H}) \, \hat{\boldsymbol{s}} \, \partial_{\varepsilon}G^{+} \, (\hat{\boldsymbol{j}} \cdot \boldsymbol{\mathcal{E}}) \right] \text{ ,}
    \label{eq:neqS_symmetry}
\end{equation}
where $\hat{\boldsymbol{s}}$, $\hat{H}$ and $\hat{\boldsymbol{j}}$ are the spin, Hamiltonian and current density operators respectively, $\mathit{f}$ is the Fermi-Dirac distribution, and $G^{+} = \text{lim}_{\eta \rightarrow 0} [\hat{H} - \varepsilon + i\eta]^{-1}$ represents the retarded Green's function. We numerically compute the Kubo-Bastin formula employing a kernel polynomial method expansion which includes the choice of a finite broadening $\eta=25 \,\text{meV}$ \cite{fan_linear_2021}. To discern the symmetry-allowed torque contributions, we compute the non-equilibrium spin density in a set of 18 magnetization directions, for each material \cite{supmat}. Note that each of these systems requires its own \textit{ab initio} and Wannier calculations as well; thus, we highlight the computational capability of the developed workflow. The exchange coupling $J_\text{ex}^{}$ is calculated as the average spectral difference between the spin majority and spin minority density of states.

\textit{SOT enhancement in Janus systems}.--- 
The equilibrium electronic structure analysis of the systems foresees an enhanced SOT response in Janus systems compared to those of the electric-field-assisted CrTe\textsubscript{2}. This is further confirmed and quantified by employing quantum transport simulations.

\begin{figure}[t]
    \centering
    \includegraphics[width=\columnwidth]{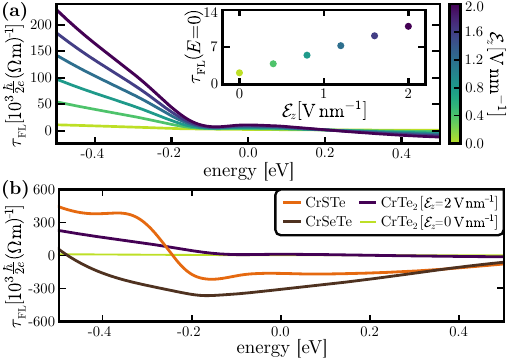}
    \caption{\textbf{(a)} Field-like spin-torque conductivity $\tau_\text{FL}$ in CrTe\textsubscript{2} computed as a function of the transversal electric field $\mathcal{E}_z$. Inset: $\tau_\text{FL}$ at the Fermi level, exhibiting a linear dependence with $\mathcal{E}_z$. \textbf{(b)} $\tau_\text{FL}$ in Janus CrXTe systems, showing much larger SOT than non-Janus CrTe\textsubscript{2}.}
    \label{fig:SOT_CrTe2_Janus}
\end{figure}

A SOT response is activated in CrTe\textsubscript{2} by applying a transversal electric field $\mathcal{E}_z$, which breaks inversion symmetry. Indeed, our results show that while the SOT response is negligible in the centrosymmetric system, all SOT components are enhanced with an increasing applied field, as showcased for the field-like torque in Fig.~\ref{fig:SOT_CrTe2_Janus}-(a). The enhancement is more prominent on the hole side of the spectrum, which can be associated with larger Fermi contours close to the K-point of the Brillouin zone. 

The field-like spin-torque conductivity at the Fermi level compares moderately to that of other two-dimensional systems, with values of the order of $10^3 \frac{\hbar}{2e} ( \Omega \text{m} )^{-1}$ \cite{hidding_spin-orbit_2020}. At a fixed energy, the spin-torque conductivity exhibits a linear dependence with respect to the applied symmetry-breaking field $\mathcal{E}_z$, as shown in the inset of Fig.~\ref{fig:SOT_CrTe2_Janus}-(a). We note that the signal for the centrosymmetric system at $\mathcal{E}_z = 0$ is non-zero, which derives from numerical approximations performed in the Wannierization procedure, yet it is negligible compared to the $\mathcal{E}_z \neq 0$ signals. The linear dependence persists through the entire $\mathcal{E}_z$ range explored, showing that even large symmetry-breaking applied fields up to $2 \,\text{V} \,\text{nm}^{-1}$ remain perturbative with respect to the internal centrosymmetric crystal field.

Janus CrXTe systems allow us to fully achieve the potential of chromium-based TMDs for SOT. The strong internal electric field generated by the asymmetric crystal structure enables a field-like torque at the Fermi level of $\sim 10^5 \frac{\hbar}{2e} (\Omega \text{m})^{-1}$ in Janus systems, 10--100 times larger than the electric-field-assisted CrTe\textsubscript{2}, as shown in Fig.~\ref{fig:SOT_CrTe2_Janus}-(b). This huge SOT enhancement is present throughout a wide energy window about the Fermi level. The obtained values are comparable with the highest torques reported in two-dimensional systems, which however rely on spin transfer from a SOC material to a ferromagnet, thus being highly susceptible to the interface quality \cite{wang_room_2017, li_spin-momentum_2018, husain_large_2020, xu_high_2020}.

\begin{figure}[t]
    \centering
    \includegraphics[width=\columnwidth]{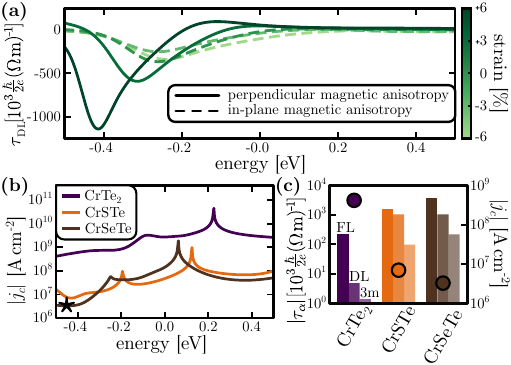}
    \caption{\textbf{(a)} 
    $\tau_\text{DL}$ in CrSTe for various strain values, exhibiting PMA for strain larger than 1\% (solid curves), and in-plane anisotropy otherwise (dashed curves)~\cite{supmat}. \textbf{(b)} Critical switching current $j_\text{c}$ in Janus CrXTe (+6\% strain), and in non-Janus CrTe\textsubscript{2} (0\% strain, and $\mathcal{E}_z=2 \,\text{V} \,\text{nm}^{-1}$). The star marks the overall optimal switching current $j_\text{c}^{\star} = 3 \times 10^6 \,\text{A} \,\text{cm}^{-2}$. \textbf{(c)} Maximal SOT efficiency (left $y$-axis; bars) and switching current (right $y$-axis; filled circles) in Janus and non-Janus systems.}
    \label{fig:SOT_strain_current}
\end{figure}

The SOT enhancement due to structural inversion symmetry breaking comes at the expense of a modification of the ferromagnetic properties. Our \textit{ab initio} calculations show that the ferromagnetic phase shifts to an in-plane magnetic anisotropy for the Janus CrXTe systems. A PMA is recovered by applying tensile strain, as shown in Fig.~\ref{fig:SOT_strain_current}-(a). Additionally, a prominent enhancement of the SOTs is observed due to strain. Indeed, the strain-induced magnetic anisotropy shift in CrXTe is driven by the Te atom, which presents larger SOC than its S/Se counterpart \cite{cui_strain-tunable_2020, liu_enhanced_2023}, thus enhancing the SOT response.

\textit{Field-free PMA switching}.--- 
We finally showcase the enormous potential of chromium-based Janus TMDs for SOT applications by computing the critical PMA switching current. The critical switching current is estimated as \cite{taniguchi_critical_2015, krizakova_spin-orbit_2022}
\begin{equation}
    j_\text{c} = \frac{M_\text{s} \, t_\text{FM} \, \sigma}{\tau_\text{DL}} \sqrt{\frac{\alpha}{\beta (2 + \alpha \beta)}} \sqrt{2B_\text{PMA}^2 - B_x^2} \text{ .}
    \label{eq:current}
\end{equation}
Here, $M_\text{s}$ is the saturation magnetization, $t_\text{FM}$ the system's thickness, $\sigma$ the longitudinal conductivity and $\beta = \tau_\text{FL} / \tau_\text{DL}$; with all of these quantities obtained from our \textit{ab initio} and quantum transport results.
For the remaining parameters; namely, the Gilbert damping $\alpha = 0.01$, and the perpendicular anisotropy field $B_\text{PMA} = 0.1 \,\text{T}$, we choose reasonable constant values, noting that they are fairly tunable by experimental conditions~\cite{qiu_electrically_2023,lee_modulating_2021}.
Finally, $B_x$ represents an in-plane effective magnetic field that drives the system out of the in-plane stationary state along $\hat{\boldsymbol{z}} \times \boldsymbol{\mathcal{E}}$ (promoted by the field-like and damping-like torques), allowing PMA switching. In conventional systems, an applied magnetic field $B_x$ is required \cite{miron_perpendicular_2011}; however, the 3m torque serves this purpose in CrXTe~\cite{liu_symmetry-dependent_2021}. 

Because $j_\text{c}$ gathers multiple magnetic and transport properties, it serves as an ultimate SOT figure of merit, providing a direct grasp of the power efficiency gain. The potential of Janus CrXTe systems is once again manifested, achieving critical switching currents 10--100 times smaller than those of non-Janus CrTe\textsubscript{2}, as shown in Fig.~\ref{fig:SOT_strain_current}-(b). The reduction of the switching current is indeed the result of enhanced SOTs in the Janus systems, evidenced in Fig.~\ref{fig:SOT_strain_current}-(c), which shows $\tau_\text{FL}$, $\tau_\text{DL}$ and $\tau_\text{3m}$ at the optimal $j_\text{c}$ energy value for each system. 
Moreover, we note that the critical switching currents calculated by Eq.~\eqref{eq:current} correspond to an upper bound for the real values, as it is derived within a macrospin approximation, whereas experimental evidence indicates that the switching occurs via domain wall nucleation and propagation \cite{baumgartner_spatially_2017, liu_symmetry-dependent_2021}.
We find an overall optimal switching current of $j_\text{c}^\star = 3 \times 10^6 \,\text{A} \,\text{cm}^{-2}$, occurring for CrSeTe with +6\% strain at $0.4 \,\text{eV}$ below the Fermi level. Experimentally, such strain can appear naturally by the substrate or the growth conditions~\cite{supmat}. This value is already highly competitive among 2D magnetic materials \cite{alghamdi_highly_2019, ostwal_efficient_2020, wang_current-driven_2019, wang_magnetic_2022}, while experimental conditions may result in further reduction of the critical current.

\textit{Conclusions}.--- 
We have found that magnetic chromium-based Janus TMDs offer a remarkable SOT performance. 
Concatenating \textit{ab initio} and quantum transport methodologies we show that the large SOT response stems from huge internal electric fields due to their asymmetric structure, yielding a competitive switching current with the additional advantage of neither requiring assistance of external fields nor the transmission of spin current through an imperfect interface.  
Such results put forward magnetic Janus TMDs as efficient materials for designing ultimate SOT-MRAM technologies.

\begin{acknowledgments}
This work was supported by the FLAG-ERA project MNEMOSYN.
J.M.D., S.R. and J.H.G. acknowledge grant PCI2021-122035-2A-2 funded by MICIU/AEI/10.13039/501100011033 and European Union “NextGenerationEU/PRTR” and the support from Departament de Recerca i Universitats de la Generalitat de Catalunya.
J.M.D. acknowledges support from MICIU (grant FPI PRE2021-097031). 
Parts of the calculations used the allocation of computational resources from GENCI–IDRIS (Grant No. 2024-A0150912036).
This project has received funding from the \textit{European Union's Horizon 2020 research and innovation programme} under grant agreement No \textit{800945} — NUMERICS — H2020-MSCA-COFUND-2017.

% The authors acknowledge funding from Ministerio de Ciencia e Innovacion (MCIN) under grant PID2022-138283NB-I00/MCIN/AEI/10.13039/501100011033 and the European Regional Development Fund.   
% S.R and J.H.G, acknowledge grant PCI2021-122035-2A-2 funded by MCIN/AEI/10.13039/501100011033 and European Union ``NextGenerationEU/PRTR” and the support from Departament de Recerca i Universitats de la Generalitat de Catalunya. 
% ICN2 is funded by the CERCA Programme/Generalitat de Catalunya and supported by the Severo Ochoa Centres of Excellence programme, Grant CEX2021-001214-S, funded by MCIN/AEI/10.13039.501100011033.  This work is also supported by MICIN with European funds‐NextGenerationEU (PRTR‐C17.I1) and by Generalitat de Catalunya.
% This project has received funding from the \textit{European Union's Horizon 2020 research and innovation programme} under grant agreement No \textit{800945} — NUMERICS — H2020-MSCA-COFUND-2017.
\end{acknowledgments}

\bibliography{bib.bib}

\end{document}

\title{Supplemental Material for ``Giant Spin-Orbit Torque in Cr-based Janus Transition Metal Dichalcogenides''} 

\author{Libor Voj\'{a}\v{c}ek}
\thanks{Authors contributed equally}
\affiliation{Universit\'{e} Grenoble Alpes, CEA, CNRS, IRIG-Spintec, 38000 Grenoble, France}

\author{Joaqu{\'i}n Medina Due{\~n}as}
\thanks{Authors contributed equally}
\affiliation{ICN2 --- Catalan Institute of Nanoscience and Nanotechnology, CSIC and BIST, Campus UAB, Bellaterra, 08193 Barcelona, Spain}
\affiliation{Department of Physics, Universitat Aut{\`o}noma de Barcelona (UAB), Campus UAB, Bellaterra, 08193 Barcelona, Spain}

\author{Jing Li}
\affiliation{Universit\'{e} Grenoble Alpes, CEA, Leti, F-38054, Grenoble, France}

\author{Fatima Ibrahim}
\affiliation{Universit\'{e} Grenoble Alpes, CEA, CNRS, IRIG-Spintec, 38000 Grenoble, France}

\author{Aurélien Manchon}
\affiliation{Aix-Marseille Université, CNRS, CINAM, Marseille 13288, France}

\author{Stephan Roche}
\affiliation{ICN2 --- Catalan Institute of Nanoscience and Nanotechnology, CSIC and BIST,
Campus UAB, Bellaterra, 08193 Barcelona, Spain}
\affiliation{ICREA --- Instituci\'o Catalana de Recerca i Estudis Avan\c{c}ats, 08010 Barcelona, Spain}

\author{Mairbek Chshiev}
\affiliation{Universit\'{e} Grenoble Alpes, CEA, CNRS, IRIG-Spintec, 38000 Grenoble, France}
\affiliation{Institut Universitaire de France, 75231 Paris, France}

\author{Jos{\'e} H. Garc{\'i}a}
\affiliation{ICN2 --- Catalan Institute of Nanoscience and Nanotechnology, CSIC and BIST, Campus UAB, Bellaterra, 08193 Barcelona, Spain}

\date{\today}

\maketitle

\onecolumngrid

% \section{Torques as a function of the in-plane lattice parameter}

% For lattice parameters slightly larger than the DFT-calculated ground state, the large peaks in all the (field-like, 3m, damping-like) torque components lying below $E_\mathrm{F}$ shift much closer to the Fermi level as shown in Fig.~\ref{fig:3m_strain_CrXTe_supplementary}.

%  \begin{figure}
%     \centering
% 	\includegraphics[width=1\textwidth]{supmat/3m_strain_CrXTe_supplementary.jpg}
% 	\caption{The conductivity and torque components as a function of in-plane lattice parameter (in-plane biaxial strain).}
% 	\label{fig:3m_strain_CrXTe_supplementary}
% \end{figure}

\section{Ab initio calculation details}

% \subsection{Short version - for main text}
% The \textit{ab initio} ground state was calculated within DFT+$U$ as implemented in \texttt{VASP}~\cite{kresse_ab_1993, kresse_efficiency_1996} with GGA-PBE pseudopotentials~\cite{perdew_generalized_1996-1} and an effective Hubbard $U_\mathrm{eff} = 3.0$~eV. The plane-wave basis was truncated at 330~eV and the Brillouin zone was sampled by a $19 \times 19 \times 1$ $\Gamma$-centered mesh. Forces were minimized below $10^{-3}$~eV/$\mathrm{\AA}$ and the total energy below $10^{-7}$~eV. The Wannier tight-binding models were constructed using the \texttt{Wannier90}~package with a 22-orbital basis (10 Cr-d and $2 \times 6$ chalcogen p-orbitals) and the interactions limited by a $25 \times 25$ real-space supercell. A real-space spin operator was derived for the Wannier models. See the Supplemental material [] for details.

\textbf{The \textit{ab initio} ground state} calculations were performed using density functional theory (DFT) as implemented in the Vienna \textit{ab initio} simulation package (\texttt{VASP}) \cite{kresse_ab_1993, kresse_efficiency_1996} with the GGA-PBE exchange-correlation functional~\cite{perdew_generalized_1996-1}. The Hubbard $U$ correction in Dudarev's formulation~\cite{dudarev_electron-energy-loss_1998} was applied to the Cr d orbitals with an effective Hubbard correction $U_\mathrm{eff} = 3.0$~eV, a value that makes the band structure closely match the experimentally-measured one~\cite{zhang_room-temperature_2021}. A Cr pseudopotential with semicore p electrons was chosen and an energy cutoff of 330~eV was used for the plane-wave basis. Brillouin zone was sampled with a 19x19x1 $\Gamma$-centered mesh. The van~der~Waals interaction was approximated by the DFT-D3 method~\cite{grimme_consistent_2010} with the Becke-Johnson damping~\cite{grimme_effect_2011}. Forces were minimized below $10^{-3}$~eV/$\mathrm{\AA}$ during the full structural relaxation and total energy below $10^{-7}$~eV to achieve the electronic self-consistency. Spin-orbit coupling was included except in the relaxation step.\\

\textbf{The Wannier tight-binding models} were constructed using the \texttt{Wannier90}~package, with a 22-orbital basis (10 Cr-d and $2 \times 6$ chalcogen-p orbitals) and a large $25 \times 25$ real-space supercell that limits the real-space interactions. A smaller supercell might very well be sufficient, see Sec.~\ref{subsec:Wannier_quality_supercell_convergence}. The \textit{Wannier} energy window was chosen to tightly encompass the 22 bands with the Cr-d and chalcogen-p character. The \textit{frozen} energy window has an upper limit to avoid including an intersecting parasitic s/dz$^2$ band and keeps an additional margin of 0.2~eV from this band, a value chosen to optimize the RMSE quality of the CrTe$_2$ Wannier model (see Sec.~\ref{sec:Wannier_quality}). Disentanglement and maximal localization were performed with 200 and 100 maximum steps, respectively. The spin operator in the maximally-localized Wannier function (MLWF) basis was derived, as described below.

\subsection{The spin operator in real space}\label{subsec:Wannier_spin_operator}

% (1) get the operator in the Wannier gauge (see Eg. 16 in Ryoo 2019 PRB or for instance Eq. 30 in Qiao 2018 PRB)
Accessing the spin expectation values of the interpolated band structure with Wannier tight-binding models requires deriving the Wannier real-space spin operator.   

% \subsection{For maximally-localized Wannier functions (MLWF)}\label{subsec:Wannier_spin_operator}

% The preliminary to obtaining the spin operator in the Wannier basis is the wannierization~\cite{marzari_maximally_2012} procedure with the \texttt{Wannier90} package~\cite{mostofi_updated_2014}, where the essential output is the semi-unitary $J \times \mathcal{J}_\textbf{q}$ matrix $V_{mn}(\textbf{q})$, where $ \mathcal{J}_\textbf{q} \geq J$ is the number of Bloch states at a specific k-point $\textbf{q}$ of the \textit{ab initio} grid being mixed together to produce $J$ number of Wannier orbitals.   

Just like any other operator that can be obtained from the \textit{ab initio} calculation (for all the \textit{ab initio} k-points $\bm{q}$), the \textit{spin operator} in the basis of \textit{ab initio} eigenstates $\mathcal{S}_{mn}^\mathrm{H}(\textbf{q})$ can be converted into its real-space representation by first applying the semi-unitary matrix $V_{m^{\prime} m}(\textbf{q})$ to convert $\mathcal{S}$ from the \textit{Hamiltonian} gauge to the \textit{Wannier} gauge

\begin{equation}
    \mathcal{S}_{m n}^{\mathrm{W}}(\bm{q})=\sum_{m^{\prime} n^{\prime}} V_{m^{\prime} m}^\dagger(\bm{q}) \cdot \mathcal{S}_{m^{\prime} n^{\prime}}^\mathrm{H}(\textbf{q}) \cdot V_{n^{\prime} n}(\bm{q}),
\end{equation}

where the spin operator $\hat{\mathcal{S}}$ projected onto the basis of the \textit{ab initio} eigenstates $\psi$ in its matrix form $\mathcal{S}_{m^{\prime} n^{\prime}}^\mathrm{H}(\textbf{q}) = \left\langle\psi_{m^{\prime} \bm{q}}|\hat{\mathcal{S}}| \psi_{n^{\prime} \bm{q}}\right\rangle$ is obtained with the help of the \texttt{vaspspn} module of \texttt{WannierBerri}~\cite{tsirkin_high_2021}. Primed indices run over the (larger) space of disentanglement bands.

Follows the direct Fourier sum over the \textit{ab initio} grid

\begin{equation}
    \mathcal{S}_{m n}(\bm{R}) \equiv \frac{1}{N_{\bm{q}}} \sum_{\bm{q}}  \mathcal{S}_{m n}^{\mathrm{W}}(\bm{q}) \cdot e^{-i \bm{q} \cdot \bm{R}} \,,
\end{equation}

with $ N_{\bm{q}} $ the number of \textit{ab initio} grid points $\bm{q}$.

Having $\mathcal{S}_{m n}(\bm{R})$ at hand, its interpolation $\overline{\mathcal{S}}_{mn}^\mathrm{H} (\bm{k})$ to an \textit{arbitrary} k-vector $\bm{k}$ involves an inverse Fourier sum $\overline{\mathcal{S}}_{mn}^\mathrm{W} (\bm{k}) = \sum_\textbf{R} \mathcal{S}_{mn} (\bm{R}) \cdot  e^{i \bm{k} \cdot \bm{R}}$ over the real-space lattice vectors $\bm{R}$ followed by a rotation back to the Hamiltonian gauge of the original eigenstates $\overline{\mathcal{S}}_{mn}^\mathrm{H} (\bm{k}) = (U^\dagger \cdot \mathcal{S}^\mathrm{W} \cdot U)_{mn}$, where $U_{mn}$ is a unitary matrix which diagonalizes the interpolated Hamiltonian $\overline{\mathcal{H}}_{mn}^\mathrm{W} (\bm{k}) = \sum_\textbf{R} \mathcal{H}_{mn} (\bm{R}) \cdot  e^{i \bm{k} \cdot \bm{R}}$.

%\subsection{For projection-only Wannier functions}

%Should be only a diagonal matrix of 2x2 sub-blocks of Pauli matrices
%As used for instance in Ref.~[cite Smaili 2021]

\section{Electrostatics of C\MakeLowercase{r}XT\MakeLowercase{e}}
The field-like torque is directly linked to the Rashba effect attributed to an inversion symmetry breaking, which comes from an electrical dipolar field across the monolayer plane.

\subsection{Internal E-fields in CrXTe monolayers}
\begin{figure}
    \centering
        % Answer: [trim={left bottom right top},clip]
	\includegraphics[trim={0 0 20cm 11.2cm},clip,width=0.5\columnwidth]{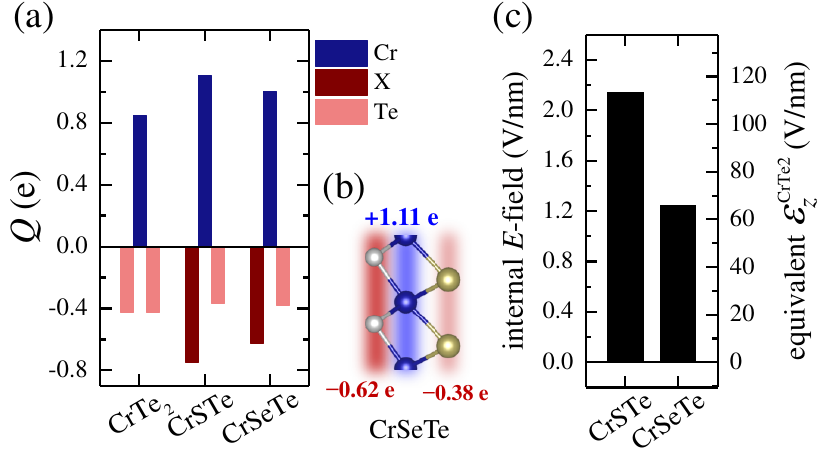}
	\caption{Electrostatics of CrXTe monolayers. (a,b) The (Bader) charge transfer $Q$ is different at the X and Te sides, giving rise to (c) a large internal electric field $\sim 2$~V/nm. The equivalent out-of-plane E-field over CrTe\textsubscript{2} would have to be an enormous $\epsilon_\mathrm{CrTe2} \, 2~\mathrm{V/nm} \sim 100$~V/nm, demonstrating the potency of CrXTe.}
	\label{fig:internal_CrXY_field}
\end{figure}

The spontaneous internal E-fields in the Janus CrXTe come from the charge imbalance, as shown in Fig.~\ref{fig:internal_CrXY_field}. This creates a work function difference and an internal electric field $\sim$~1--2~V/nm in CrXTe. We calculate these values from the work function difference $\Delta \phi$ [Fig.~\ref{fig:CrTe2_relative_permittivity}(a)] divided by the thickness of the layer, which is derived from the positions of the chalcogen atom centers extended by their Wigner-Seitz radii. 

\subsection{Relative permittivity of CrTe\textsubscript{2} monolayer}

\begin{figure}
    \centering
	\includegraphics[trim={0 0 14cm 8.7cm},clip,width=4.3in]{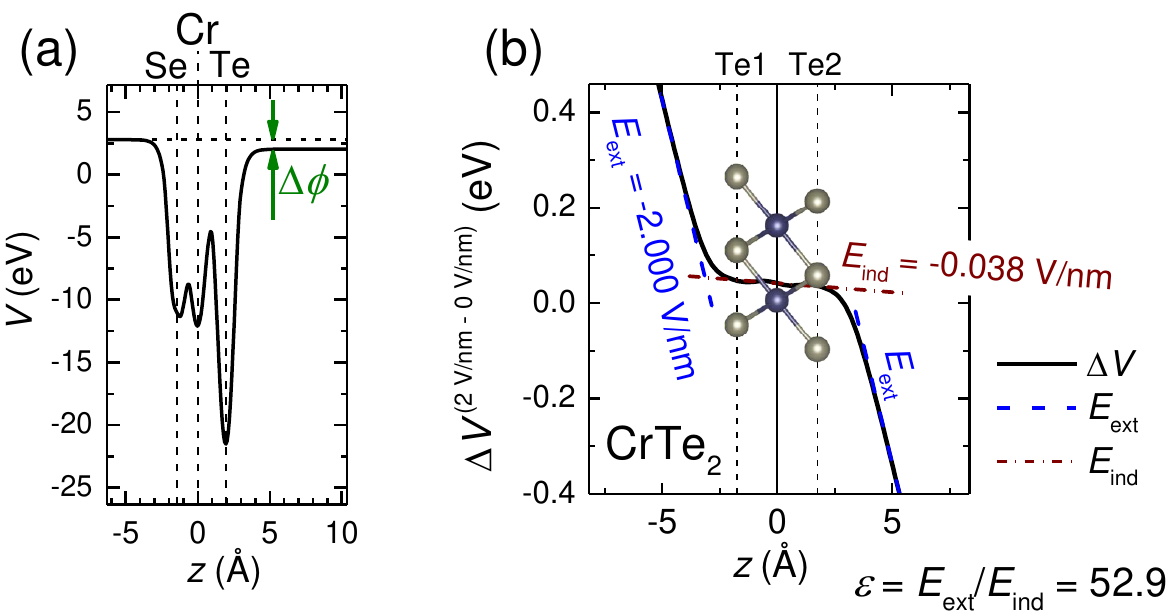}
	\caption{Electrostatics of CrXTe. (a) The ($xy$-averaged) potential in CrSeTe is different at the Se and Te sides. This work function difference translates into an internal electric dipole. (b) The relative permittivity of CrTe$_2$ monolayer can be calculated from the difference between the $xy$-averaged potential of CrTe$_2$ monolayer under an external $E_\mathrm{ext} = 2$~V/nm and without an external E-field. The potential gradient corresponds to the E-field magnitude. The ratio of the E-field outside (in vacuum) and inside the layer corresponds to $\epsilon_\mathrm{CrTe2}$.}
	\label{fig:CrTe2_relative_permittivity}
\end{figure}

Subtracting the $xy$-averaged potential of the CrTe\textsubscript{2} monolayer under an out-of-plane E-field of 2~V/nm and under no E-field, we can estimate its relative permittivity from the gradient of the potential inside the layer and in vacuum, see Fig.~\ref{fig:CrTe2_relative_permittivity}. In vacuum, the potential gradient corresponds to the imposed $E_\mathrm{ext} = 2$~V/nm, which is largely screened inside CrTe$_2$ to $E_\mathrm{ind} = 0.038$~V/nm. The $E_\mathrm{ind} / E_\mathrm{ext}$ ratio gives the relative permittivity of CrTe$_2$, $\epsilon_\mathrm{CrTe2} = 52.9.$ This is a prefactor that should be applied to the \textit{internal} electric fields of CrXTe 

\begin{equation}
    \mathcal{E}_\mathrm{ext}^\mathrm{equivalent \, for \, CrTe2} = \epsilon_\mathrm{CrTe2} \, \mathcal{E}_\mathrm{int}^\mathrm{CrXTe}
\end{equation}

to get an equivalent \textit{external} E-field over CrTe\textsubscript{2} to achieve a comparable Rashba splitting.

\subsection{The equivalent external E-field to the Janus internal E-field}

In Tab.~\ref{tab:CrXY_vs_CrX2}, the ratio between $\chi_\mathrm{FL}$ of CrXY and gated CrTe$_2$ gives an estimation of the required $E_\mathrm{ext}$ which should be applied to CrTe$_2$ to obtain a similar $\chi_\mathrm{FL}$ in CrTe$_2$ as are spontaneously present in CrXY. The values are colossal 30 to 60~V/nm. They correspond well to a simple multiplication of the CrXY internal E-fields $E_\mathrm{int}$ with the relative permittivity of CrTe$_2$ $\epsilon_\mathrm{CrTe2} = 52.9$, giving a 113 and 66~V/nm. 

% Table generated by Excel2LaTeX from sheet 'Sheet7'
\setlength{\tabcolsep}{6pt}
\begin{table}[htbp]
  \centering
  \caption{The calculated internal E-fields $E_\mathrm{int}$ in CrXY monolayers and an estimated equivalent external field $E_\mathrm{ext}^\mathrm{CrTe2} \simeq \epsilon_\mathrm{CrTe2} \cdot E_\mathrm{int}$ that should act on CrTe$_2$ to reach a similar effect. Further, the calculated field-like torques $\chi_\mathrm{FL}$ in Janus CrXY monolayers and an estimated external E-field that should act on CrTe$_2$ to reach similar torque magnitudes. Both these estimations result in huge $E_\mathrm{ext}^\mathrm{CrTe2} \approx 30$ to 50~V/nm, illustrating the superiority of Janus CrXY for producing the Rashba effect and field-like torques.}
  \begin{threeparttable}
    \begin{tabular}{lrrrr}
    \hline
    \multicolumn{1}{c}{} & \multicolumn{1}{c}{$E_\mathrm{int}$} & \multicolumn{1}{c}{$E_\mathrm{ext}^\mathrm{CrTe2} \simeq \epsilon_\mathrm{CrTe2} \cdot E_\mathrm{int} $} & \multicolumn{1}{c}{$\chi_\mathrm{FL}$}  & \multicolumn{1}{c}{$E_\mathrm{ext}^\mathrm{CrTe2} \simeq 2.0 \cdot \chi_\mathrm{FL} / \chi_\mathrm{FL}^\mathrm{CrTe2 \, @ \, 2.0\,V/nm}$} \\
    \multicolumn{1}{r}{} & \multicolumn{1}{c}{V/nm} & \multicolumn{1}{c}{V/nm}  & \multicolumn{1}{c}{10$^3 \hbar$/2e (Ohm m)$^{-1}$} & \multicolumn{1}{c}{V/nm} \\
    \hline
    CrSTe & 2.14   & \textbf{113} &163    & \textbf{30} \\
    CrSeTe & 1.25   & \textbf{66} & 308   & \textbf{57} \\
    \hline
    \end{tabular}%
    \begin{tablenotes}
        \item $ \epsilon_\mathrm{CrTe2} = 52.9; \; \; \; \chi_\mathrm{FL}^\mathrm{CrTe2 \, @ \, 2.0\,V/nm} = 10.9$~10$^3 \hbar$/2e (Ohm m)$^{-1}$
    \end{tablenotes}
  \end{threeparttable}
  \label{tab:CrXY_vs_CrX2}%
\end{table}%

\section{Strain (lattice parameter variation)}\label{sec:strain}

\begin{figure}
    \centering
	\includegraphics[width=0.9\textwidth]{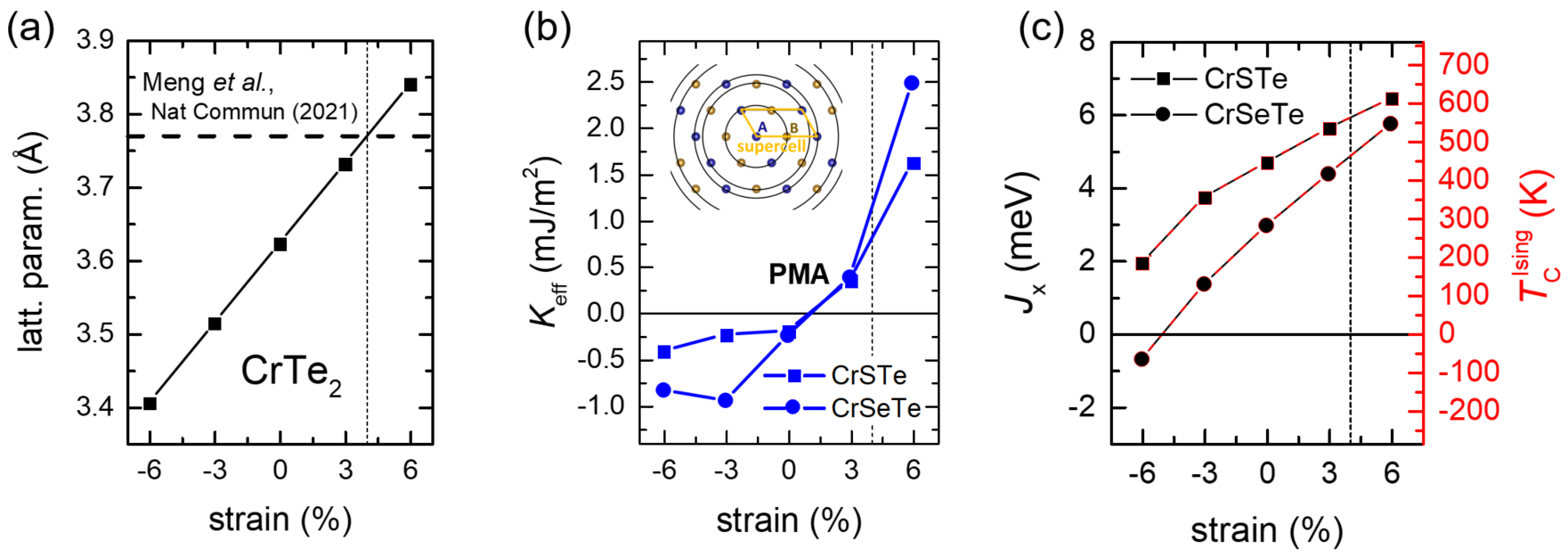}
	\caption{The magnetic anisotropy and exchange coupling with lattice parameter variation ("strain"). 0\% corresponds to the \textit{ab initio}-predicted lattice constant. (a) The experimental lattice parameter of CrTe$_2$~\cite{meng_anomalous_2021} is $\approx 4\%$ larger than the \textit{ab initio} prediction. Both (b) the effective uniaxial anisotropy and (c) the nearest-neighbor exchange interaction (and the consequent Ising Curie temperature) largely increase for higher lattice parameters in CrXTe. The vertical dashed line corresponds to the $\approx 4\%$ latt. param. difference that might be expected from the comparison of DFT and experiment of CrTe$_2$ from (a). Inset of (b) shows the 2x1x1 supercell used for the Heisenberg exchange parameter calculation.}
	\label{fig:strain}
\end{figure}

Although theoretical studies suggest that the in-plane lattice parameter of CrTe$_2$ is reduced in its monolayer form compared to bulk~\cite{liu_structural_2022}, experimental measurements~\cite{zhang_room-temperature_2021, meng_anomalous_2021} suggest that ultrathin films of CrTe$_2$ retain their lattice constant reasonably close to bulk. The experimentally measured lattice parameter of ultrathin CrTe$_2$ films $a = 3.77 \, \mathrm{\AA}$~\cite{meng_anomalous_2021} is $\approx 4\%$ larger than our calculated monolayer value, as shown in Fig.~\ref{fig:strain}(a). Since Janus CrXTe monolayers have not yet been fabricated and their experimental structure may also slightly differ from our \textit{ab initio} predictions, we deem it necessary to provide our results as a function of \textit{strain}, which is not necessarily a physical strain, but rather a variation of the lattice parameter with respect to the predicted \textit{ab initio} value.

As shown in Fig.~\ref{fig:strain}(b-c), the consequence of such \textit{positive} strain is an increase in both the effective magnetic anisotropy and the Heisenberg exchange coupling, linked to the Curie temperature.

The appropriate effective magnetic anisotropy in case of 2D magnets~\cite{lado_origin_2017}

\begin{equation}
    K_\mathrm{eff} = K_\mathrm{u} + Z (J_{z} - J_{x}) + E_\mathrm{demag}
\end{equation}

is a sum of the calculated single-ion anisotropy $K_\mathrm{u}$, the (super)exchange interaction anisotropy $Z(J_{z} - J_{x})$ with the number of nearest neighbors $Z = 6$ and the demagnetizing energy $E_\mathrm{demag} = -\mu_0 M_\mathrm{s}^2 / 2$ with its small value of $\approx -0.07$ and $-0.06$~mJ/m$^2$ for CrSTe and CrSeTe respectively. 

Note that the dominant contribution is expected to be the anisotropic superexchange $Z(J_{z} - J_{x})$, while the single-ion contribution $K_\mathrm{u}$ should be negligible~\cite{lado_origin_2017}.

All these quantities can be calculated from the \textit{ab initio} total energies, considering the XXZ Hamiltonian~\cite{lado_origin_2017}

\begin{equation}\label{eq:XXZ_Hamiltonian}
    H= -K_u \sum_i\left(S_i^z\right)^2 - \frac{J_{x}}{2} \underset{i \neq j}{\sum \sum}\left(S_i^x S_j^x+S_i^y S_j^y\right)- \frac{J_{z}}{2} \underset{i \neq j}{\sum \sum}\left(S_i^z S_j^z\right)
\end{equation}

where $i$ runs over all Cr atoms and $j$ runs over all of their 6 nearest neighbors (thereby double counting is present and redeemed by the factor $\frac{1}{2}$ coming with $J$). Constructing a 2x1x1 supercell [see the inset of Fig.~\ref{fig:strain}(b)] of the CrXTe unit cells with both ferromagnetic (FM) and antiferromagnetic (AF) configurations along both the $x$ and $z$ directions, it follows from Eq.~\ref{eq:XXZ_Hamiltonian} that the \textit{ab initio} ground state energies are $E^{\mathrm{FM}}_x = E_0 -6S^2 J_x$,  $E^{\mathrm{AF}}_x = E_0 + 2S^2 J_x$,  $E^{\mathrm{FM}}_z = E_0 -2S^2 K_\mathrm{u} - 6S^2 J_z$ and $E^{\mathrm{AF}}_z = E_0 -2S^2 K_\mathrm{u} + 2S^2 J_z$, from which

\begin{equation}
\begin{aligned}
        J_{x(z)} &=\left(-E^{\mathrm{FM}}_{x(z)} + E^{\mathrm{AF}}_{x(z)}\right) / 8 S^2 \,, \\
    K_\mathrm{u} &=\left( E^{\mathrm{FM}}_x + 3 E^{\mathrm{AF}}_x - E^{\mathrm{FM}}_z - 3 E^{\mathrm{AF}}_z \right) / 8 S^2 \,,
\end{aligned}
\end{equation} 

where $S = 3/2$ is the magnitude of the Cr spin~\cite{lado_origin_2017}.

In Fig.~\ref{fig:strain}(c) we also plot the strain-enhanced Curie temperature in the Ising limit~\cite{torelli_calculating_2018}

\begin{equation}\label{eq:Curie_Ising}
    T_\mathrm{C}^\mathrm{Ising} = \frac{S^2 \tilde{T_\mathrm{c}}}{k_\mathrm{B}}  J_{x} \,, 
\end{equation}

where the constant dimensionless $\tilde{T_\mathrm{c}} = 3.64$ for a hexagonal lattice~\cite{torelli_calculating_2018}, $k_\mathrm{B}$ is the Boltzmann constant.

\section{Quality of the Wannier models}\label{sec:Wannier_quality}

We provide a quantitative estimate of the quality of the derived Wannier tight-binding models, which aim to accurately describe the \textit{ab initio} band energies and spin expectation values.
For this, the band structure along the $\mathrm{K-}\Gamma\mathrm{-M-K}$ with 101 points along each segment is calculated both from \textit{ab initio} and the derived Wannier TB model and compared to obtain the RMSE errors as

\begin{equation}
    \mathrm{RMSE}(E) = \sqrt{\frac{1}{N}  \sum_\mathrm{\langle kpath \rangle}(E^\mathrm{DFT}-E^\mathrm{WannTB})^2}
\end{equation}
and analogically for the spin expectation values. $N$ is the number of summed data points. The averaging is performed inside a $E_\mathrm{F} \pm 0.5$~eV energy window.

\subsection{Quality convergence with the Wannier supercell size}

\begin{figure}[h]
    \centering
    % [trim={left bottom right top},clip]
	\includegraphics[width=0.8\textwidth, trim={0 0 0 1.8cm}, clip]{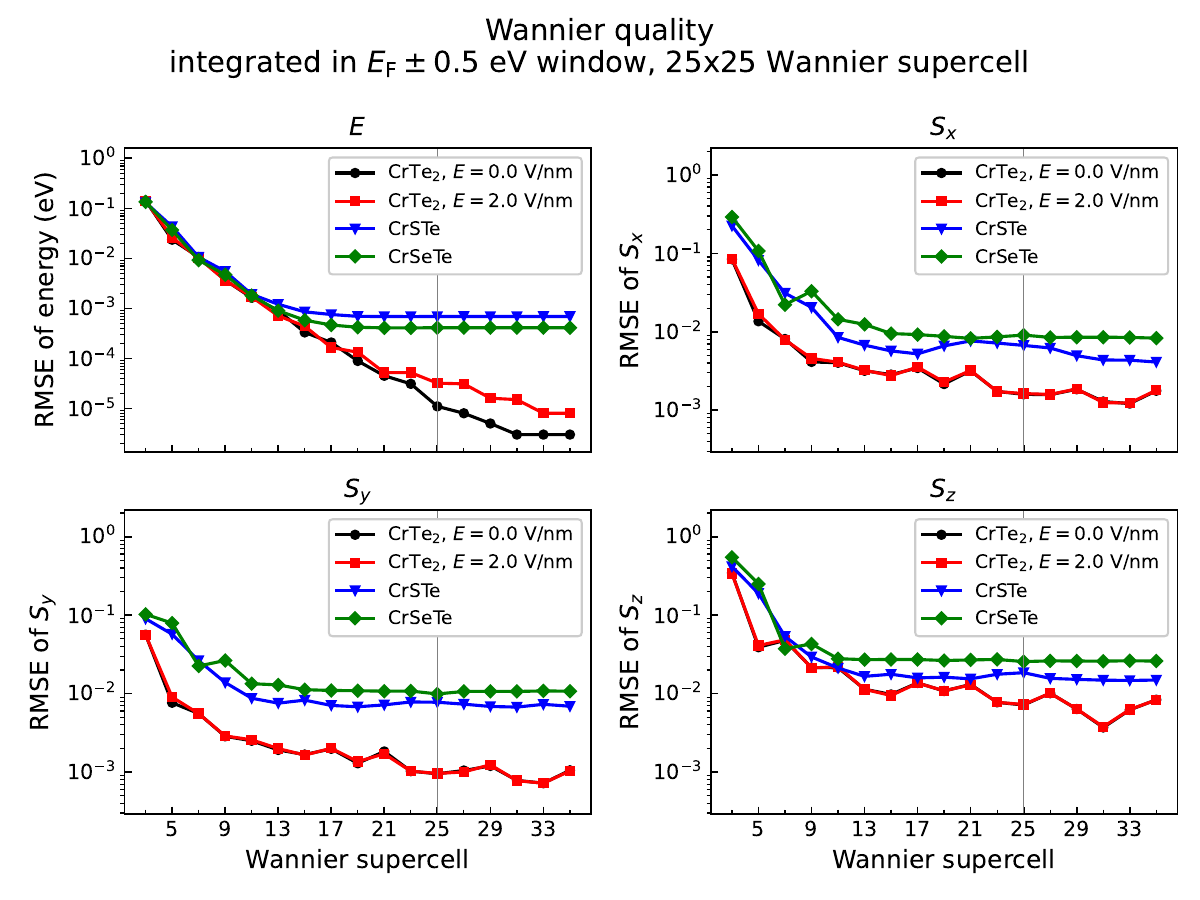}
	\caption{The root mean square error of energy and spin expectation values of the Wannier TB model vs. the original DFT calculation, averaged along the standard $\mathrm{K-}\Gamma\mathrm{-M-K}$ \textit{k}-path and within $\pm$~0.5~eV window around the Fermi level. The RMSE of energy exponentially decreases with increasing Wannier supercell (interaction cutoff). The vertical line denotes the $25 \times 25$ supercell, used for further transport calculations.}
	\label{fig:Wannier_quality_vs_Wannier_supercell}
\end{figure}

% script batch_SWEEP_WANNIER_KPOINTS_quality_with_spn_file_part.job.sh
The real-space interactions need to be limited by a large enough supercell to converge the RMSE below a chosen threshold, as plotted in Fig.~\ref{fig:Wannier_quality_vs_Wannier_supercell}. A $25 \times 25$ supercell provides sufficient accuracy. The saturation of the \textit{RMSE of energy} for CrXTe indicates a (small) constant offset of the whole energy spectrum due to slightly ($\approx 1$~meV) misaligned Fermi energy values. 

% slide 141 in "CrXY monolayers - SPINTEC with ICN2 part 2.pptx"

\subsection{Quality of the final models}\label{subsec:Wannier_quality_supercell_convergence}

\begin{figure}
    \centering
	\includegraphics[width=0.8\textwidth, trim={0 0 0 1.8cm}, clip]{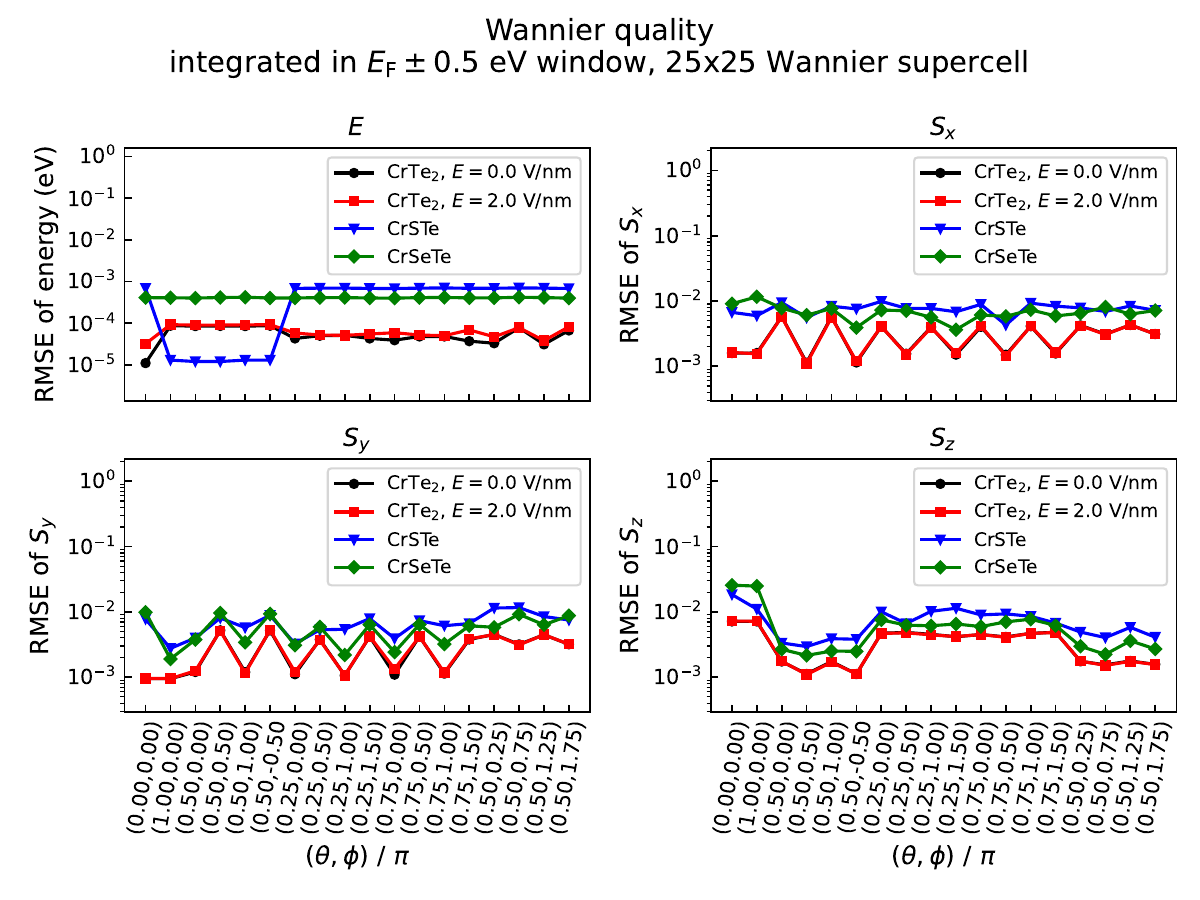}
	\caption{The root mean square error of energy and spin expectation values of the Wannier TB model vs. the original DFT calculation, averaged along the standard $\mathrm{K-}\Gamma\mathrm{-M-K}$ \textit{k}-path and within $\pm$~0.5~eV window around the Fermi level. The azimuthal and polar angles $\theta$ and $\phi$ denote calculations with different magnetization directions. The errors are small: around 1 meV in energy and 0.01 in spin expectation values.}
	\label{fig:Wannier_quality}
\end{figure}

 Figure~\ref{fig:Wannier_quality} shows the RMSE of energy and spin for CrTe$_2$ (with and without external $E$-field), CrSTe, and  CrSeTe calculated with different \textit{ab initio} magnetization directions $\Vec{m} = (\sin\theta\cos\phi, \, \sin\theta\sin\phi, \, \sin\theta)$. The errors are fairly small: $\approx 1$~meV for energy and $\approx 0.01$ for spin expectation values.

\section{Non-Equilibrium Spin Density and Symmetry-Allowed Contributions}

Consistent with the linear transport regime, we expand the non-equilibrium spin density $\boldsymbol{S}$ to linear order in terms of its $\mathcal{E}_j$ ($j=x,y$) dependence. On the other hand, since the magnetization direction does not constitute a perturbative parameter, we expand $\boldsymbol{S}$ up to second order with respect to $\hat{\boldsymbol{m}}$. 
We express the non-equilibrium spin density as $\boldsymbol{S} = \boldsymbol{S}_{0} + \boldsymbol{S}_\text{3m}$, where the first term groups the torque contributions which are allowed in arbitrary non-centrosymmetric systems, while the second represents torque contributions enabled by the reduced 3m symmetry. The first contribution reads
\begin{equation}
\begin{split}
    \boldsymbol{S}_0 =& \, \chi_\text{FL}^{} [1 - \xi_\text{FL}^{}(m_x^2 + m_y^2)] \hat{\boldsymbol{z}} \times \boldsymbol{\mathcal{E}} - \chi_\text{DL}^{} \hat{\boldsymbol{m}} \times (\hat{\boldsymbol{z}} \times \boldsymbol{\mathcal{E}}) - \chi_\text{DL}^{z} (\hat{\boldsymbol{m}} \cdot \boldsymbol{\mathcal{E}}) \hat{\boldsymbol{z}} - \chi_\parallel^{} [\hat{\boldsymbol{m}} \cdot (\hat{\boldsymbol{z}} \times \boldsymbol{\mathcal{E}})] \hat{\boldsymbol{m}} - \chi_\parallel^{z} [\hat{\boldsymbol{m}} \cdot (\hat{\boldsymbol{z}} \times \boldsymbol{\mathcal{E}})]m_z \hat{\boldsymbol{z}} \text{ ,}
\end{split}
\end{equation}
showing that all systems allow for many contributions beyond the conventional field-like and damping-like torques \cite{jmedina2024}, respectively represented in the terms proportional to $\chi_\text{FL}^{}$ and $\chi_\text{DL}^{}$, where the field-like torque is modulated by the total in-plane magnetization via $\xi_\text{FL}^{}$. $\chi_\text{DL}^{z}$ accounts for an anisotropic damping of the in-plane and out-of-plane magnetization components towards the steady state along $\hat{\boldsymbol{z}} \times \boldsymbol{\mathcal{E}}$.
$\chi_\parallel^{}$ represents a non-equilibrium spin density component parallel to the magnetization which cannot exert torque, while $\chi_\parallel^{z}$ corresponds to an out-of-plane anisotropy of the aforementioned component which, along with the field-like torque, induces an elliptic precession of the magnetization about $\hat{\boldsymbol{z}} \times \boldsymbol{\mathcal{E}}$. A set of unconventional torques are additionally enabled by the system's 3m symmetry, which are represented in
\begin{equation}
\begin{split}
    \boldsymbol{S}_\text{3m}
    =& \,\chi_\text{3m}^{} [(\mathcal{E}_x m_y + \mathcal{E}_y m_x) \hat{\boldsymbol{x}} + (\mathcal{E}_x m_x - \mathcal{E}_y m_y) \hat{\boldsymbol{y}}] 
    -\chi_\text{3m}^{(2)} m_z [ (-\mathcal{E}_x m_x + \mathcal{E}_y m_y) \hat{\boldsymbol{x}} + (\mathcal{E}_x m_y + \mathcal{E}_y m_x) \hat{\boldsymbol{y}}] \\&
    +  \chi_\text{3m}^{z} (\mathcal{E}_x m_x^2 - \mathcal{E}_x m_y^2 - 2 \mathcal{E}_y m_x m_y) \hat{\boldsymbol{z}} \text{ .}
\end{split}
\label{eq:S_3m}
\end{equation}
The term proportional to $\chi_\text{3m}$ represents the so-called 3m torque, while the other second-order terms stem from a contribution of the form $\hat{\boldsymbol{m}} \times \boldsymbol{s}_\text{3m}^{}$, with $\boldsymbol{s}_\text{3m}^{}$ the aforementioned first order 3m spin density. The 3m torque derives from a current-induced in-plane magnetic anisotropy which, when applying a driving electric field $\boldsymbol{\mathcal{E}} = \mathcal{E}_x \hat{\boldsymbol{x}}$ perpendicular to the mirror plane, lies along the $m_x=-m_y$ axis \cite{johansen_current_2019}. Thus, the first order 3m torque $\chi_\text{3m}^{}$ accounts for precession about the anisotropy axis, while the second order torques $\chi_\text{3m}^{(2)}$ and $\chi_\text{3m}^{z}$ account for a damping parallel to it.

\section{Transport simulations: complete results}

\subsection{Strainless}
\begin{figure}
    \centering
    \includegraphics[width=\textwidth]{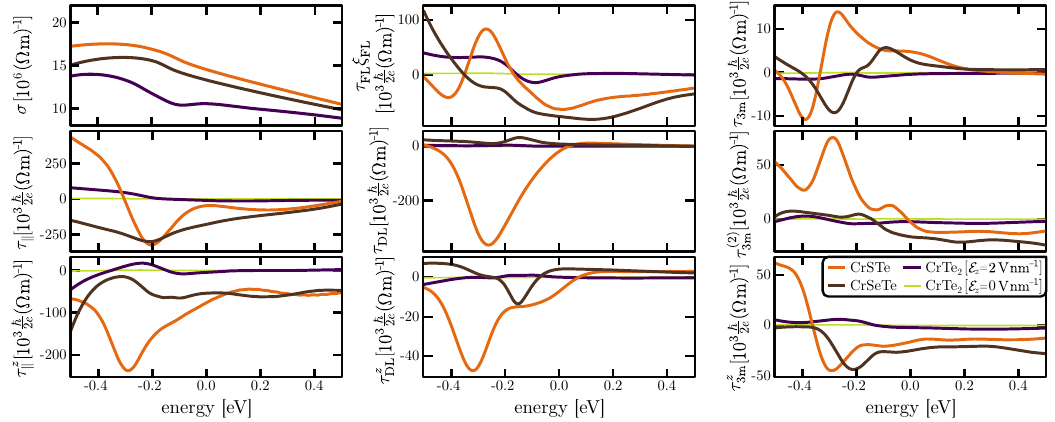}
    \caption{Longitudinal conductivity $\sigma$, in units of $10^6 (\Omega \,\text{m})^{-1}$, and all spin-torque conductivities $\tau_\alpha$, in units of $10^3 \frac{\hbar}{2e} ( \Omega \,\text{m} )^{-1}$, for all systems. The longitudinal conductivity of CrTe\textsubscript{2} is unaffected by $\mathcal{E}_z$}
    \label{fig:SOT_all}
\end{figure}

We present the complete spin-orbit torque results for all the studied systems, at 0\% strain. In order to disentangle the symmetry-allowed torque contributions we compute the non-equilibrium spin density for a set of 18 magnetization directions for each system, consisting of 8 evenly spaced $\varphi$ points for $\theta=\pi/2$, 4 evenly spaced $\varphi$ points for $\theta=\pi/4, 3\pi/4$, and the two poles $\theta=0,\pi$, with ($\theta,\varphi$) the magnetization polar and azimuthal angles. The spin-torque conductivities are shown in Fig.~\ref{fig:SOT_all}. We observe that all torque contributions are much larger in the Janus structures than in the electric field assisted CrTe\textsubscript{2}, while they are vanishing in centrosymmetric CrTe\textsubscript{2}. Additionally, the results for the longitudinal conductivity show that it remains of the same order in Janus vs non-Janus systems.

\subsection{Strain-dependent}
\begin{figure}
    \centering
    \includegraphics[width=\textwidth]{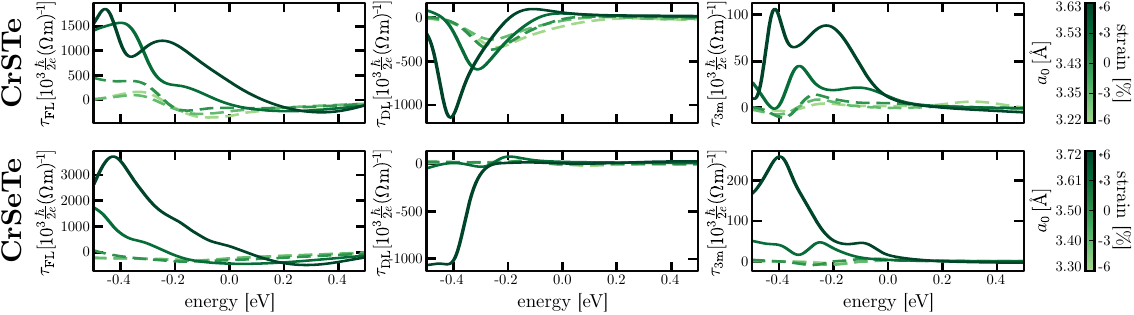}
    \caption{Field-like, damping-like and 3m spin-torque conductivities, in units of $10^3 \frac{\hbar}{2e} ( \Omega \,\text{m} )^{-1}$ as a function of strain. Solid and dashed curves correspond to perpendicular and in-plane magnetic anisotropy respectively. The corresponding lattice parameter $a_0$ is indicated in the colorbar.}
    \label{fig:SOT_strain}
\end{figure}

Fig.~\ref{fig:SOT_strain} shows the field-like, damping-like and 3m torques as a function of strain. All torques are enhanced with strain, retrieving a perpendicular magnetic anisotropy for strain $\geq 1\%$.

\newpage

\bibliography{bib.bib}